\documentclass[aps,prb,preprint,onecolumn,amsmath,amssymb,superscriptaddress,eqsecnum,floatfix,scrartcl]{revtex4-1}
\pdfoutput=1      
\usepackage{amssymb}  
\usepackage{color,soul} 
\usepackage{graphicx}  
\usepackage[tight]{subfigure}  
\usepackage{mathrsfs}
\usepackage[pdftex,pdfusetitle,bookmarks=true,colorlinks=true,citecolor=blue,urlcolor=blue,linkcolor=magenta]{hyperref}
\usepackage{hypcap}

\begin{document}
\title{Operator scrambling and quantum chaos}    
\author{Xiao Chen}
\email{xchen@kitp.ucsb.edu} \affiliation{Kavli Institute for
Theoretical Physics, University of California at Santa Barbara, CA
93106, USA}

\author{Tianci Zhou}
\email{tzhou13@illinois.edu} \affiliation{Department of Physics
and Institute for Condensed Matter Theory, University of Illinois
at Urbana-Champaign, IL 61801-3080, USA}

\date{\today}

\begin{abstract}
Operator scrambling is a crucial ingredient of quantum chaos. Specifically, in the quantum chaotic system, a simple operator can become increasingly complicated under unitary time evolution. This can be diagnosed by various measures such as square of the commutator (out-of-time-ordered correlator), operator entanglement entropy {\rm etc}. In this paper, we discuss operator scrambling in three representative models: a chaotic spin-$1/2$ chain with spatially local interactions, a 2-local spin model and the quantum linear map. In the first two examples, although the speeds of scrambling are quite different, a simple Pauli spin operator can eventually approach a ``highly entangled" operator with operator entanglement entropy taking a volume law value (close to the Page value). Meanwhile, the spectrum of the operator reduced density matrix develops a universal spectral correlation which can be characterized by the Wishart random matrix ensemble. In the second example, we further connect the 2-local model into a one dimensional chain and briefly discuss the operator scrambling there. In contrast, in the quantum linear map, although the square of commutator can increase exponentially with time, a simple operator does not scramble but performs chaotic motion in the operator basis space determined by the classical linear map.  We show that once we modify the quantum linear map such that operator can mix in the operator basis, the operator entanglement entropy can grow and eventually saturate to its Page value,  thus making it a truly quantum chaotic model.
\end{abstract}

\maketitle

\section{Introduction}
\label{sec:intro}

Quantum chaos is an interesting phenomenon in physics and has deep connection with the random matrix theory\cite{BGS1984} and quantum thermalization\cite{Srednicki1994,Deutsch1991}. Recently, it has been proposed that the square of the commutator between two simple operators\cite{Larkin1969}, i.e., 
\begin{align}
C(t)=\langle [\hat V(t), \hat W][\hat V(t), \hat W]^\dag\rangle
\end{align}
can measure the dynamics of quantum chaos and characterizes the quantum butterfly effect\cite{Shenker2013a,Maldacena2015}. In some large $N$ systems, $C(t)$ can grow exponentially in time with the rate denoted as the quantum Lyapunov exponent, in analogy with the classical chaotic systems\cite{Maldacena2015}. The Sachdev-Ye-Kitaev (SYK) model is  one of the maximally chaotic models whose Lyapunov exponent saturates the chaos upper bound\cite{Kitaev2014, Kitaev2015, Sachdev1993,Maldacena2015, Maldacena2016}. 

The dynamics of $C(t)$ is determined by the Heisenberg operator $\hat V(t)=e^{i\hat Ht}V(0)e^{-i\hat Ht}$. Under unitary time evolution, this operator can spread out in Hilbert space and become increasingly complicated. In other words, the quantum information encoded in this operator is delocalized and this phenomenon is called scrambling. In the past decade, many  researchers are interested in designing various fast scramblers and exploring mechanisms of information scrambling in the quantum states\cite{Hayden2007,Sekino2008, Brown2012, Lashkari2013,Roberts2015b,Hosur2016,Brown2017}. In this paper, we will investigate the scrambling of operator $\hat V(t)$. In particular, we will focus on three representative models: a chaotic spin-$1/2$ chain with local interactions, a 2-local spin-$1/2$ model and a modified quantum linear map. We will show the dynamics of $\hat V(t)$ are different in these three models.

In the chaotic spin-$1/2$ chain, under time evolution, a local Pauli operator can become increasingly non-local with the speed bounded by the Lieb-Robinson velocity\cite{Roberts2015b,Nahum2017,Keyserlingk2017}. The time for this operator to spread over the entire system is linearly proportional to the system size and is denoted as the scrambling time. Due to the small onsite Hilbert space, there is no exponentially growing regime in $C(t)$. In contrast, in the 2-local Hamiltonian, the scrambling time is much shorter and scales as $\log N$, where $N$ is the number of spins. This is because the interaction exists between {\it each pair} of $1/2$ spins (with the strengths of the same order). We will argue that this fast scrambling process leads to the exponential growth of $C(t)$ at early time. The operator scrambling has also been discussed recently in the SYK model in Ref.~\onlinecite{Roberts2018}.

In both cases, although the speeds of scrambling are different, a simple Pauli operator can eventually approach a ``featureless" random operator after long time evolution, which can be considered as an operator version of Berry's conjecture on random pure state in chaotic system\cite{Berry1977}\footnote{Notice that under unitary time evolution, the eigenvalue of the Heisenberg operator is always the same}. Operator scrambling suggests that there is strong correlation between subsystem and a good measure of this correlation is the so-called operator entanglement entropy (EE)\cite{Prosen2007,Prosen2009,Dubail2017,Ho2017,Zhou2017,Xu2018,Jonay2018}. Our results indicate that in both cases, the subsystem (equal or less than half of the system) operator EE can approach the volume law which is very close to the Page value, suggesting that the information is fully scrambled in the subsystem\cite{Page1993}. The random matrix physics, which is an important indicator of quantum chaos\cite{BGS1984}, shows up as the operator becomes more complicated. Numerically, we observe the emergence of spectral correlation in the spectrum of operator reduced density matrix as operator EE grows. We find that it fully develops when the operator spreads over the entire Hilbert space and can be described by the Wishart random matrix\cite{ChenLudwig2017SpectralCorrelations,loggas}.

Furthermore, we investigate the operator dynamics in the quantum linear map\cite{Hannay1980,Kurlberg2000}. This model represents a large class of quantum mechanical models which are chaotic in the classical limit. In these models, $C(t)$ between two quantum operators which possess meaningful classical limits can increase exponentially in time. However, different from the 2-local Hamiltonian, the Lyapunov exponent here has a classical origin and may not be related with the operator scrambling. In fact, we observe that the trajectory of the operator $\hat V(t)$ in a suitably chosen basis is determined by the classical dynamics\cite{Kurlberg2000}. It does not become more complicated as time evolves. This is evident as we find that the operator EE is very small and no spectral correlation developed in the spectrum of the reduced operator density matrix $\hat\rho_A^{\hat V}(t)$. 


We further study the quantum linear map perturbed by nonlinear shear\cite{Dematos1995}. We notice that at early time, the quantum operator moves in the basis space (determined by the classical dynamics), accompanied by a weak mixing in the basis operators. This mixing (or superposition) will lead to the operator scrambling. After sufficient time evolution, the operator is fully scrambled with operator EE saturating to the Page value. Notice that in the modified quantum linear map, there are two different time scales. The first one is the Lyapunov time $t_L$, after which the quantum correction becomes important and $C(t)$ cannot be approximated by the classical dynamics anymore\cite{Cotler2017,Rozenbaum2017,Tian2004}. The other one is the scrambling time, which is determined by the strength of the perturbation and can be much longer than the Lyapunov time.

Before going into the detail, we summarize the main results of these three models in Table~\ref{summary_model}. The rest of the paper is organized as follows: In Sec.~\ref{sec:local_int}, we investigate the operator scrambling in chaotic spin-$1/2$ chain model and demonstrate that it can be characterized by both the operator entanglement entropy and entanglement spectrum. In Sec.~\ref{sec:2-local}, we explore the operator scrambling in 2-local spin-$1/2$ Hamiltonian and derive the dynamics of  ``typical height" of the operator. We further show that this leads to the exponential growth of the square of commutator. In Sec.~\ref{dot_chain}, we use this 2-local spin Hamiltonian to build up a one dimensional model and study the possible temporal-spatial dynamics in it. In Sec.~\ref{sec:linear_map}, we discuss the dynamics of Heisenberg operator in the quantum linear map. In addition, we investigate the operator scrambling in the modified quantum linear map and generalize the results to other similar quantum mechanical models. We conclude in Sec.~\ref{sec:conclusion} with some final remarks.






%
%

\begin{table}[htbp]
\centering
\begin{tabular}{c|c|c|c}
 &  Lyapunov regime &Scrambling time & Spectral correlation \\ \hline
spin-$\frac{1}{2}$ chain with local interaction& Nonexistent & $L/v_B$& Yes \\ \hline
2-local Hamiltonian & $\Delta t\log N$ & $\Delta t\log N$& Yes \\ \hline
modified quantum linear map & $\log K/\lambda_+$ & $f(\kappa)\log K$ & $\begin{matrix} \kappa=0,\ \mbox{No}\\ \kappa>0,\ \mbox{Yes} \end{matrix}$
\end{tabular}
\caption{The main results of three different models, where Lyapunov regime denotes the time regime in which $C(t)$ can grow exponentially in time.}
\label{summary_model}
\end{table}

\section{chaotic spin-$1/2$ chain}
\label{sec:local_int}

In this section, we consider a generic chaotic spin-$1/2$ chain with local interaction and study the scrambling of an operator $\hat{O}(x, t)$ initially localized at position $x$. $\hat O(x,t=0)$ is local in the sense that an observer outside its support will not see any change when it acts on a state\footnote{In other words, the operator is the tensor products of the on-site identity operators beyond its support.}. As time evolves, this operator will become increasingly non-local and at zeroth order the operator will grow ballistically on its bilateral ends no faster than the Lieb-Robinson velocity\cite{Lieb1972}. 

However, the operator end is not sharply defined and we should actually study the operator length probability distribution. To clarify this concept, we need to first take an operator basis $\{\hat{\mathcal{B}}_j\}$, which itself has a well-defined length. Such basis in spin-$1/2$ chain can be taken as the tensor products of the one-site Pauli matrices and identity. Then the length of the basis is naturally the largest distance between two one-site Pauli matrices\footnote{A one-site Pauli matrix has length 1. An identity operator has size zero. }. $\hat{O}(x, t)$ is generally a superposition in this Pauli string basis
\begin{align}
\label{eq:expand-pauli}
\hat O(x,t)=\sum_j\alpha_j(t)\hat{\mathcal{B}}_j. 
\end{align}
where the coefficients $|\alpha_j(t)|^2$ as probability are normalized to 1: $\sum_j\alpha_j(t)^2=1$. The length distribution inside $\hat{O}(x, t)$ is given by the probability of length-$l$ basis operator at time $t$
\begin{equation}
f( l, t ) = \sum_{j } |\alpha_j(t)|^2  \delta(\text{length } ( \hat{\mathcal{B}}_j) = l ) 
\end{equation}
The zeroth order solution $f(l, t) = \delta( l  - v_B t )$ neglects the possible dispersion of the wavepacket as it moves with group velocity $v_B$. 

\begin{figure}
\centering
\includegraphics[width=.5\textwidth]{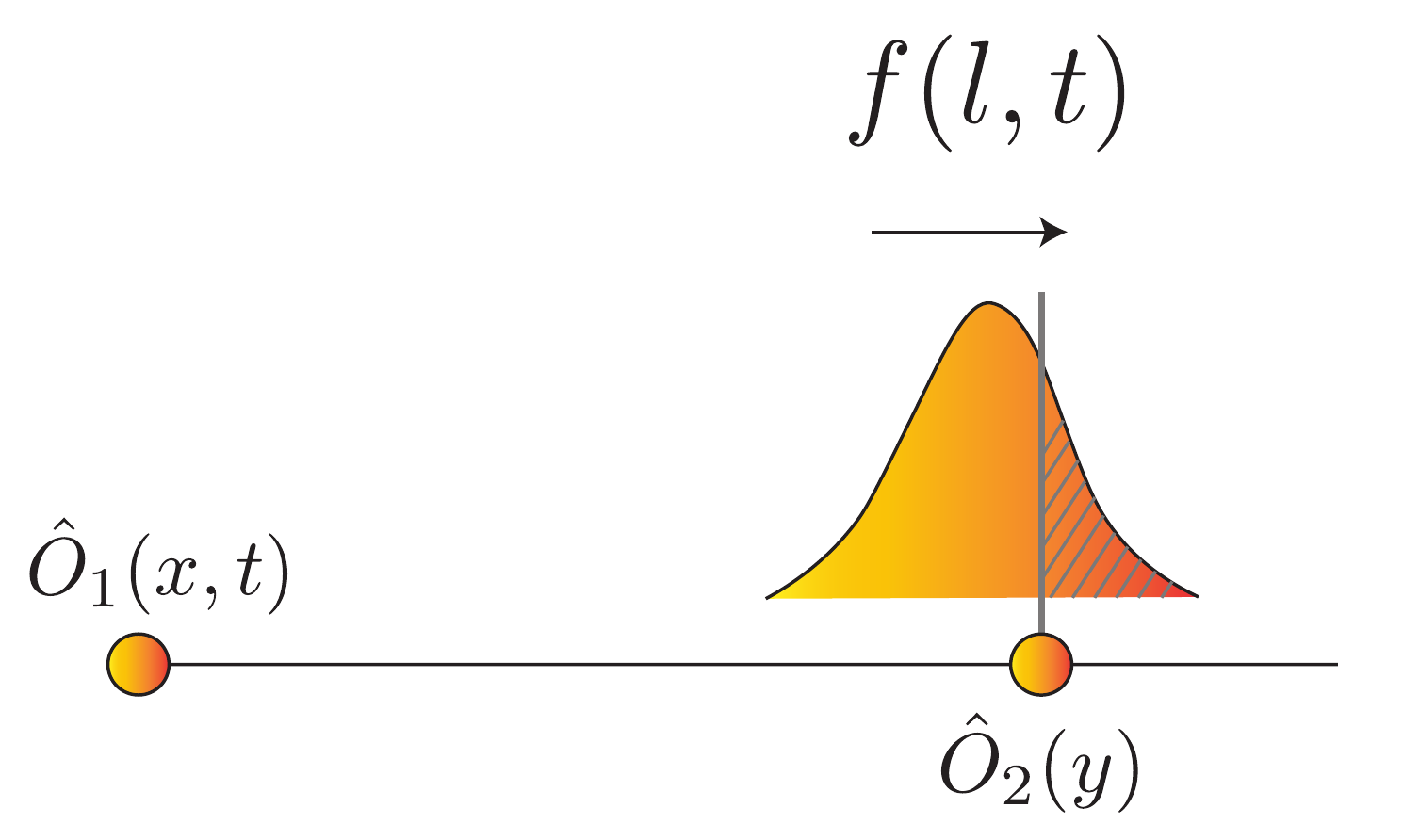}
\caption{The schematic of length distribution function $f(l,t)$ of $\hat O_1(x,t)$. The shape of $f(l,t)$ can be measured by the square of commutator between $\hat O_1(x,t)$ and $\hat O_2(y)$.}
\label{fig:schematic_op_scramble}
\end{figure}

Although exact solutions are generally not available, recently random unitary circuits provide simple tractable examples of the distribution function $f(l,t)$\cite{Nahum2017,Keyserlingk2017,Rakovszky2017,Khemani2017}. The underlying rule is that the end of each basis will perform an independent biased random walk under the evolution of the random gate. Therefore in operator $\hat{O}(x, t)$, an ensemble of such basis will lead to a diffusive broadening of $f(l,t)$. To avoid the issue of the bilateral growth, we put the initial operator on the left boundary of the chain, and then the length is equal to the location of the operator's right end point. For the random circuit, $f(l,t)$ is a moving Gaussian with its width scales as $\sqrt{t}$\cite{Nahum2017,Keyserlingk2017}. This diffusive wave front picture is further numerically verified in some chaotic spin-1/2 chain model\cite{Keyserlingk2017}. Notably, such description is not restricted to the spatially local interactions. When the interaction range is increased, the wave front can be less localized. For example, a power law decay interaction (with proper exponent) gives a moving wavepacket with the front having a near-exponential shape\cite{chen_measuring_2017}.

The length distribution is one aspect about the operator growth: the extension in the spatial direction. Another aspect is how the operator mixes in its already occupied territories. For example a long operator concentrated on a single basis (the Pauli basis above) can be very long, but is still localized and not ``thermalized" in the operator space. This motivates us to study the operator reduced density matrix $\hat\rho_A^{\hat{O}}(t)$, which is the reduced density matrix if we treat an operator as a doubled sided state\cite{Hosur2016}. In particular, we will throw away the information of the basis and study the basis-independent eigenvalues of $\hat\rho_A^{\hat{O}}(t)$. 

\begin{figure}
\centering
\includegraphics[width=.5\textwidth]{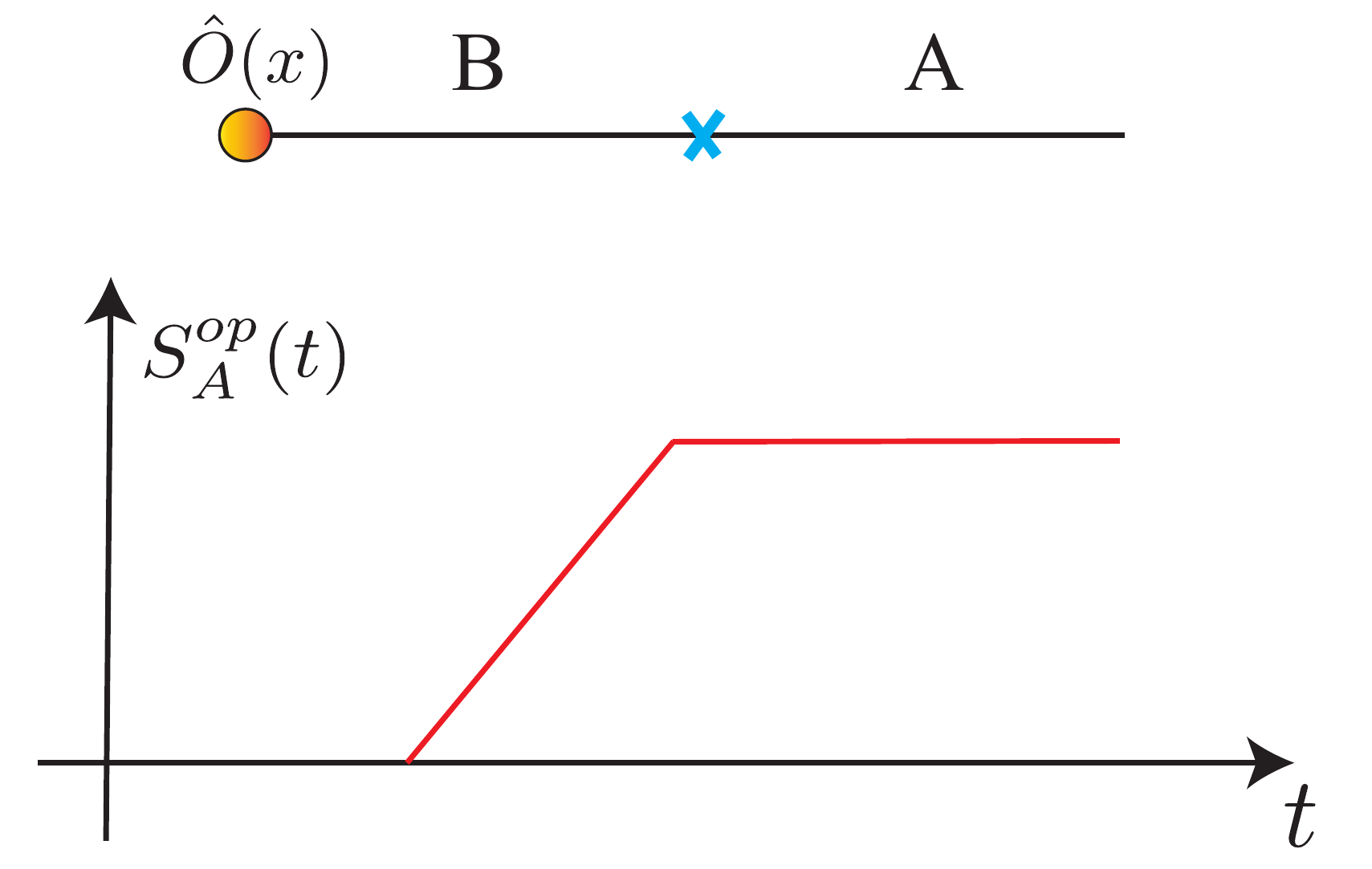}
\caption{The schematic of $S_A^{op}(t)$ of an initial local operator $\hat O(x)$.}
\label{fig:schematic_op_EE}
\end{figure}

One such measure is the operator entanglement entropy of $\hat\rho_A^{\hat{O}}(t)$, which reflects how fast the operator is scrambled. In particular, we spatially bipartition the spin chain into two half chains A and B and explore the time evolution of $\hat\rho_A^{\hat{O}}(t)$ of an initially local operator $\hat O(x, t=0)$.  We assume that $\hat O(x, t = 0)$ is in regime A at $t=0$, and the
operator EE is zero (See Fig.~\ref{fig:schematic_op_EE}). As time evolves, the operator expands spatially. After the operator front enters into subsystem B, the operator EE starts to grow linearly with time and eventually saturates to a constant close to the Page value $2\log |H_A|-|H_A|^2/2|H_B|^2$\cite{Page1993}, where $|H_A|$ and $|H_B|$ are the Hilbert space dimensions for A and B (we assume $|H_A|\leq |H_B|$). It is close to the Page value because we are effectively at infinite temperature. This is in clear contrast to the Clifford circuit evolution \cite{gottesman_heisenberg_1998} or the quantum linear map discussed in Sec.~\ref{sec:linear_map}, where the operator is always living on a single basis. We believe that only the former can be termed as quantum chaotic evolution, see more discussions in Sec.~\ref{sec:linear_map}. 

As operator EE grows with the time, the spectral correlation starts to develop in the eigenvalues of $\hat\rho_A^{\hat{O}}(t)$\cite{ChenLudwig-prep}, similar to the same quantity studied in the globally quenched state\cite{ChenLudwig2017SpectralCorrelations}. The spectral correlation can be conveniently characterized by the spectral form factor\cite{HaakeBook,StoeckmannBook} -- the Fourier transform of the pair correlation function between two of the eigenvalues $\lambda_i$ of $\hat\rho_A^{\hat{V}}$,
\begin{align}
g(\tau)=\langle \sum_{i,j}e^{-i\tau(\lambda_i-\lambda_j)} \rangle
\end{align}
where $\langle\cdots\rangle$ stands for averaging over spectral distribution. After sufficient time evolution, the operator is fully scrambled in the subsystem A and we observe a strictly linear growth of $g(\tau)$ as a function of $\tau$ (often denoted as the ramp\cite{Cotler2016}). This demonstrates the presence of the universal spectral correlation in $\hat\rho_A^{\hat{V}}$ and is consistent with the Wishart random matrix theory\cite{ChenLudwig2017SpectralCorrelations,loggas}. The emergence of the universal level repulsion in $\hat\rho_A^{\hat{V}}(t)$ after long time evolution is an indicator of quantum chaos. We will use this method to explore operator scrambling and quantum chaos in the other two models.

In the end, we would like to comment on the relation between the square of commutator $C(t)=-\langle[\hat O_1(x,t), O_2(y)]^2\rangle$ with the above. Our criterion for an operator to scramble in these one dimensional chain models is that operator becomes increasingly nonlocal under unitary time evolution, accompanied by the strong mixing in the operator basis. The former is reflected by the moving wavepacket in the length distribution $f(l,t)$, and the later can be detected from operator EE or the level repulsion of the operator reduced density matrix. In the case that both are satisfied, $C(t)$ between $\hat O_1(x,t)$ and $\hat O_2(y)$ probes the area of front of $f(l>y,t)$. This is because only the basis non-commutative with $\hat O_2(y)$ can contribute to $C(t)$, which lies in the shaded region of the front (See Fig.~\ref{fig:schematic_op_scramble}). In fact, by the strong mixing, the fraction of basis non-commutative with $W$ is always a constant, hence $C(t)$ is proportional to the area of the shaded wave front $\int f(x>y,t)dx$. In the spin-$1/2$ chain with power-law interaction (with fine-tuned exponent), since the front of $f(l,t)$ has a near-exponential shape, $C(t)$ can increase exponentially with time, which is absent in the model with local interaction where $f(l,t)$ is described by the Guassian front\cite{chen_measuring_2017}.

\section{2-local qubit Hamiltonian}
\label{sec:2-local}

The model we discussed in Sec.~\ref{sec:local_int} contains {\it spatially} local interactions and the scrambling time is linear to the system size. In this section, we allow any pair of the spins in the system to interact, a form of interaction usually called 2-local in the literature\cite{Kempe2006}. These types of the model are therefore strongly chaotic and believed to be fast scramblers\cite{Hayden2007,Sekino2008, Brown2012, Lashkari2013}.

We assume the interaction strengths of these 2-local terms are of the same order. As an example, consider the following Hamiltonian composed of $N$ $\frac{1}{2}$-spins\cite{erdos_phase_2014}, 
\begin{align}
H=\frac{1}{\sqrt{9N}}\sum_{1\leq i<j\leq N}\sum_{a,b=1}^3 \alpha_{a,b,(i,j)}\sigma_i^a\sigma_j^b
\end{align}
where the interaction strength $\alpha_{a,b,(i,j)}$ are independent random variables with zero mean and unit variance.

Similar to the previous model, we study the operator scrambling of an initially simple operator $\hat O(t=0)=\hat\sigma_i^a$ under the unitary time evolution. At any time, the evolved operator $\hat O (t)$ can be expanded in the Pauli string basis as in Eq.~\eqref{eq:expand-pauli} with the coefficients satisfying the normalization constraint $\sum_j\alpha_j(t)^2=1$. For each Pauli string operator $\hat{\mathcal{B}}_j$, we define the height of this operator to be the number of Pauli operators $\hat{\sigma}^{(x,y,z)}$ in this string. This naturally leads to the height probability distribution 
\begin{align}
P(h,t)=\sum_j\alpha_j(t)^2\delta(\text{height of } \hat{\mathcal{B}}_j=h)
\end{align}
subjected to normalization $\sum_{h=1}^N P(h,t)=1$.



We make the following assumptions on the height distribution:
\begin{enumerate}
\item The distribution $P(h, t)$ is concentrated on a particular value $h(t)$ at time $t$ in the large $N$ limit, i.e., the operator has a {\it typical height} $h(t)\in [0,N]$. 
\item The coefficients $\alpha_j$ are uniformly distributed among the sector of the Pauli strings with the same height $h$. 
\end{enumerate}

With the first assumption, scrambling time can be defined as the period after which the typical height  reaches an appreciable fraction of the maximal height $N$. Hence inverting the typical height function $h(t)$ gives the scrambling time. We now estimate the typical height $h(t)$ from the operator dynamics. An operator of height $h(t)$ is generated from the height $1$ operator at the beginning by increasing its height successively to $2$, $3$ until finally reaching $h(t)$. The total time required is the sum of the transition times in each step that increases the height by $1$. An important observation is that the transition time is not a constant but depends on the number of interaction terms that extends the operator. This can be understood from the short time evolution of Pauli string operator $\hat{\mathcal{B}}$ with height $l$,
\begin{equation}
\hat{\mathcal{B}}(t)\sim \hat{\mathcal{B}}(0)+i[\hat H, \hat{\mathcal{B}}(0)]t
\end{equation}
Thanks to the 2-body nature of the interaction, only the terms non-commutative with $\hat{\mathcal{B}}$ can increase its height by $1$ . The number of those terms is
\begin{align}
\mathcal{N}(l\to l+1)=l\times 2\times 3\times (N-l).
\end{align}
The transition time should be inversely proportional to the number of terms. Hence if the transition time from $l=1\to l=2$ is $\Delta t$, then the transition time from $l\to l+1$ is
\begin{align}
\Delta t\frac{\mathcal{N}(1\to 2)}{\mathcal{N}(l\to l+1)}=\Delta t \frac{N-1}{l(N-l)}
\end{align}
Therefore the time to reach a typical height $h$ is about
\begin{align}
t=\sum_{l=1}^h  \Delta t \frac{N-1}{l(N-l)}\approx \Delta t\log\frac{h(N-1)}{N-h}
\end{align}
which in turn determines the height function 
\begin{equation}
\label{eq:typical_h}
h(t)=\frac{Ne^{\frac{t}{\Delta t}}}{N+e^{\frac{t}{\Delta t}}-1}.
\end{equation}
At early time, when $e^{\frac{t}{\Delta t}}\ll N$, $h(t)$ has an exponential increase in Eq.~\eqref{eq:typical_h}. As a consequence, the scrambling time will scale as $\Delta t\log N$, which is much shorter than the linear scrambling time for the model with (spatially) local interaction.  

As a side note, the differential equation satisfied by $h(t)$ is the logistic differential equation
\begin{equation}
\frac{d h}{d (t / \Delta t)}=h(1-\frac{h}{N})
\end{equation}
describing the population growth, whose solution -- the logistic function -- has many applications in different areas\cite{Gershenfeld1999}. In the population growth model, the right hand side has a factor $h$ denoting the fertility proportional to the current population and $(1 - \frac{h}{N})$ factor as a result of the limited resources consumable from the environment. In the context of the operator scrambling, the right hand side means that the number of interactions participating in extending the operators is maximized at half of the system's height limit.

\begin{figure}
\centering
\includegraphics[width=.5\textwidth]{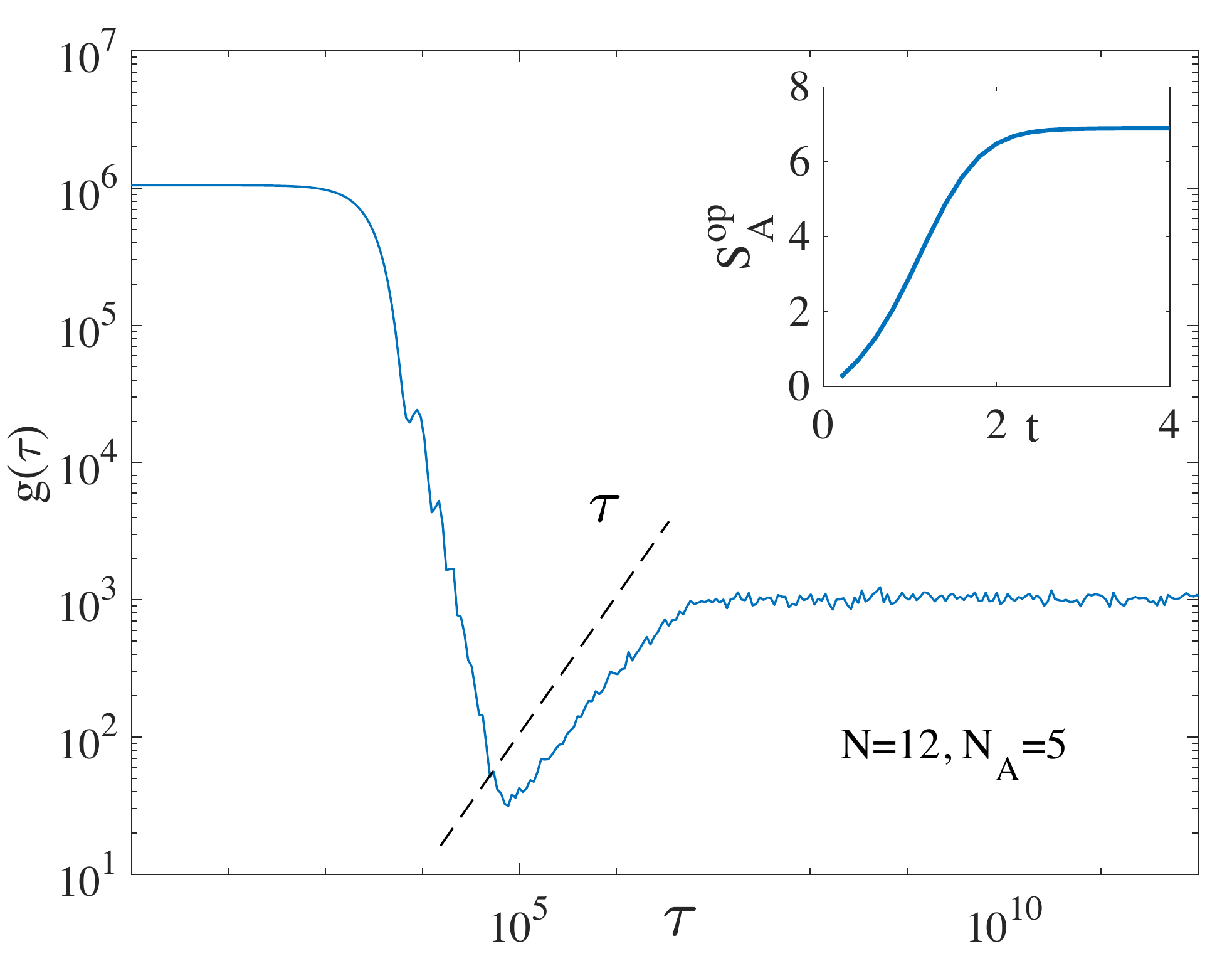}
\caption{Spectral form factor for the  $\hat \sigma^x$ operator reduced density matrix. The curves are averaged over 200 times in the interval $t\in[21,200]$. The inset is operator EE vs time.}
\label{fig:Z_2_dot}
\end{figure}

Next we consider the square of the commutator $C(t)=-\langle [\hat\sigma_i(t),\hat\sigma_i]^2\rangle $ which is proportional to the weight of those operators that are non-commutative with $V$. At time $t$, the typical height is $h(t)$, and there are $3^h{ N \choose h }$ such operators. Among them, there are $3^{h-1}{ N -1 \choose h - 1 }$ operators with height $h(t)$ that also occupy the site of $V$. According our second assumption, the weight are uniformly distributed. So $C(t)$ will be proportional to the ratio
\begin{equation}
C(t) \propto \frac{{ N -1 \choose h(t) - 1 }}{{ N \choose h(t) }} = \frac{h(t)}{N}.
\end{equation}
Therefore $C(t)$ also grows exponentially when $t< \Delta t\log N$. The growth rate $\frac{1}{\Delta t}$ is model dependent which we expect to scale linearly against the strength of the interaction.


\subsection{One-dimensional model}
\label{dot_chain}

The 2-local Hamiltonian system is effectively a ($0+1$) dimensional quantum dot for the lack of spatial locality in its interaction. One interesting generalization is to form a one dimensional lattice of the quantum dots with local inter-dot interaction. This construction is similar to the recently proposed SYK chain model\cite{Gu2017}. Here the operator scrambling occurs simultaneously in both the onsite Hilbert space and spatial direction. One possible  way to describe the evolution of the typical height $h(x,t)$ is to add a diffusion term in the logistic differential equation, i.e.,
\begin{align}
\frac{\partial h}{\partial t}=D\frac{\partial ^2h}{ \partial x^2}+\lambda h(1-\frac{h}{N})
\label{fisher}
\end{align}
This nonlinear diffusion equation is called Fisher's equation (also known as Kolmogorov-Petrovsky-Piskunov equation) where $\lambda=1/\Delta t$ and $D$ is the diffusion constant. One important feature of this equation is the existence of the traveling wave solution $h(x-vt)$ with initial condition $h(x\to -\infty)/N=1$ and $h(x\to \infty)/N=0$. The velocity is $v=c\sqrt{D\lambda}$ and stability requires the parameter $c \ge 2 $\cite{Kolmogorov1937study,Ablowitz1979}. 

\begin{figure}[hbt]
\centering
 \subfigure[]{\label{fig:FKPP_evo} \includegraphics[width=.4\textwidth]{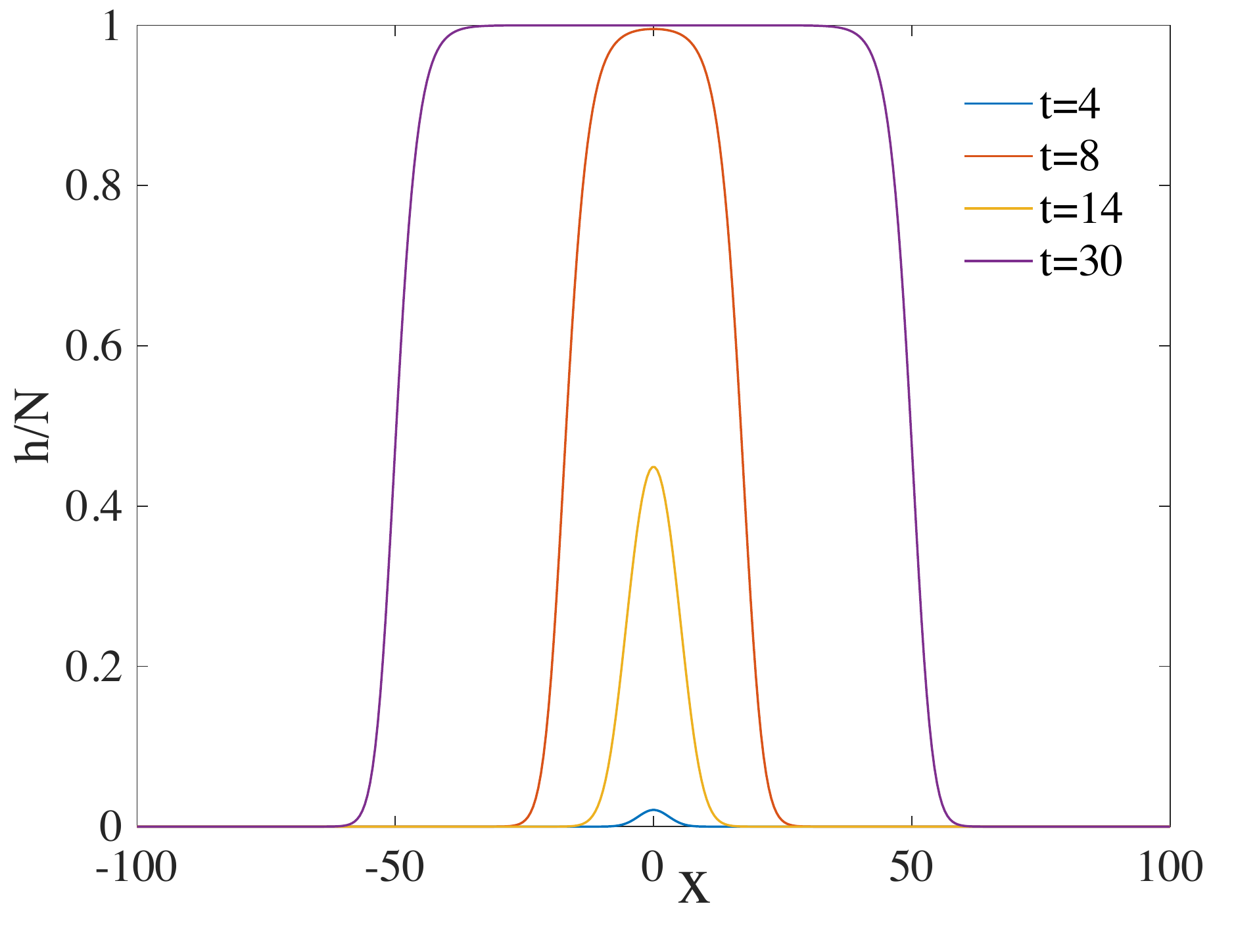}}
 \subfigure[]{\label{fig:FKPP_sol} \includegraphics[width=.4\textwidth]{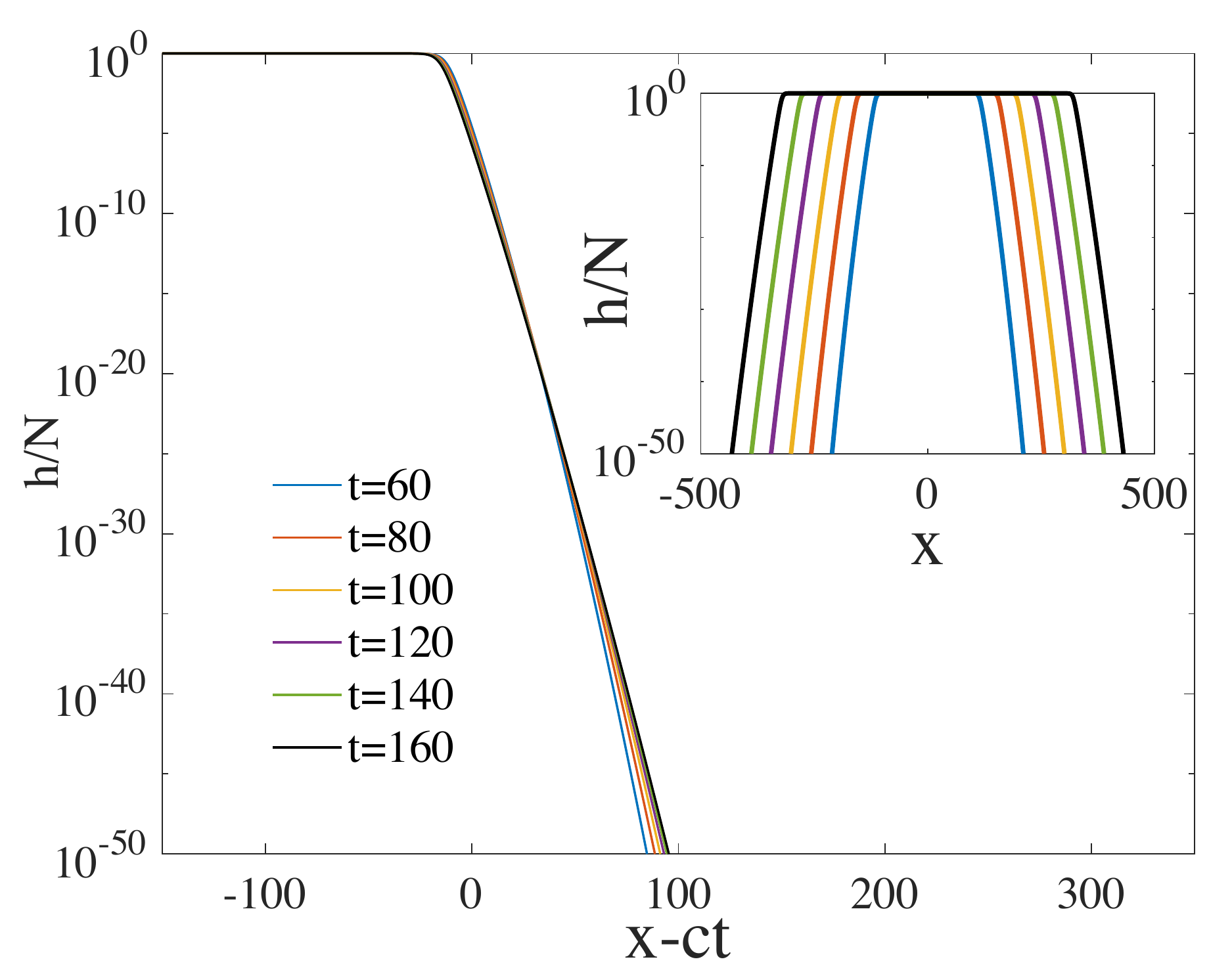}}  
\caption{
(a) The numerical simulation of Fisher's equation with $D=1$ and $\lambda=1$. The initial condition is a small Gaussian packet localized around $x=0$. (b) The $h(x,t)/N$ curves  at various $t$ (shown in the inset) are collapsed to a single curve after shifting the x-axis by $-ct$. Here we take $c=2.1$ for the right wave front, which is very close to the lower bound $c=2$.
} 
\label{fig:Fisher}
\end{figure}

Our motivation is to understand the temporal-spatial dynamics of $h(x,t)$ of an initial simple operator at some position. Therefore we choose the initial boundary condition as a small Gaussian packet localized at position $x=0$ and numerically solve Eq.\eqref{fisher}. At early time when $h(x,t)< N$, we expect
\begin{align}
\frac{h(x,t)}{N}\sim e^{\lambda t-\frac{x^2}{4 Dt}}
\label{h_1}
\end{align}
$h(x,t)$ grows exponentially with time and spreads out diffusively. At the time scale $\log N$, $h(x\approx 0,t )$  saturates to $N$ (See Fig.~\ref{fig:FKPP_evo}). The size of $h=N$ plateau increases with the time. Furthermore, we observe the appearance of a {\it stable traveling wave}  by collapsing $h(x,t)$ at various $t$ to a single curve (Fig.~\ref{fig:FKPP_sol}). Around $t-x/v=0$, the wave front takes a simple form 
\begin{align}
\frac{h(x,t)}{N}\sim e^{a\lambda(t-x/v)}
\label{h_2}
\end{align}
where we numerically find $a=0.9$.

Similar to the quantum dot model, the square of commutator $C(x,t)$ between two operators of distance $x$ is proportional to $h(x,t)$. Therefore we expect to observe a crossover of $C(t)$ from Eq.\eqref{h_1} when $t<\log N$ to Eq.\eqref{h_2} when $t\gg \log N$. Notice that Eq.\eqref{h_1} has been proposed in weakly interacting diffusively metal\cite{Patel2017} while the exponential form of Eq.\eqref{h_2} is found in holographic model and SYK chain\cite{Shenker2013a,Gu2017}. It would be interesting to have a better understanding of this crossover behavior in a microscopic model. Actually, similar crossover behavior has been discussed in an electronic system\cite{Aleiner2016}. We leave the detailed study of this part in the future.

\section{quantum linear map}
\label{sec:linear_map}

In this section, we study the operator scrambling in the quantum linear map which is an instructive quantum mechanical model with many properties exactly solvable. Before we study this model in detail, we first briefly review the classical linear map (also known as Arnold's cat maps\cite{Arnol1968ergodic}).

The classical linear map is the linear automorphism of the unit torus phase space given by
\begin{align}
\begin{pmatrix}
q\\p
\end{pmatrix}\to
\begin{pmatrix}
a & b\\ c & d
\end{pmatrix}
\begin{pmatrix}
q \\ p 
\end{pmatrix} \mbox{mod}\ 1
\label{classical_cat}
\end{align}
where the matrix $M=\left(\begin{smallmatrix} a & b\\ c & d\end{smallmatrix}\right)\in \mbox{SL}(2,\mathbb{Z})$. The determinant is equal to one so that this map is area preserving (canonical). Also it preserves the periodic boundary condition of the torus as $M$ has integer valued entries. The Lyapunov exponents $\lambda_\pm$ of the linear map are given by the logarithm of the eigenvalues of $M$\cite{Hirsch2012}. When $\mbox{Tr} M>2$, this map is hyperbolic and has $\lambda_+>0$ ( $\lambda_++\lambda_-=0$). The chaotic linear map is known to be fully ergodic and mixing\cite{Arnol1968ergodic}. In the following study, we will consider a simple case with $a=2,b=1,c=3, d = 2$ and $\lambda_{\pm}=\log(2\pm\sqrt{3})$. 

The linear map can be quantized on the square torus with finite Hilbert space\cite{Hannay1980,Kurlberg2000}. We define $|q_n\rangle$ and $|p_n\rangle$ to be position and momentum eigenstates with $n=0,1,\cdots, K-1$, where $K$ is dimension of the Hilbert space. The position and momentum translation operators are defined through $
\hat\tau|q_n\rangle=|q_{n+1}\rangle$ and $\hat\sigma|p_n\rangle=|p_{n+1}\rangle$. 
Hence $\hat{\sigma}$ and $\hat{\tau}$ can be represented as $\mathbb{Z}_K$ rotor operators,
\begin{align}
\hat{\sigma}=\begin{pmatrix}
1 & 0 & \cdots & 0  \\
0 & \omega & \cdots & 0  \\
\vdots & \vdots & \ddots & \vdots\\
0 & 0 & \cdots & \omega^{K-1}
\end{pmatrix},\quad
\hat{\tau}=\begin{pmatrix}
0 & \cdots & 0 & 1  \\
1 & \cdots & 0 & 0  \\
\vdots & \ddots & \vdots & \vdots\\
0 & \cdots & 1 & 0
\end{pmatrix}
\end{align}
where $\omega=e^{-2\pi i/K}$. $\hat{\sigma}$ and $\hat{\tau}$ satisfy $\hat\sigma^K=\hat\tau^K=1$ with the commutation relation given by $\sigma\tau=\omega \tau \sigma$. 

In the position representation, the quantum propagator for the quantum linear map with $M=\left(\begin{smallmatrix} 2 & 1\\ 3 & 2\end{smallmatrix}\right)$ is obtained by path integral method and takes this form (Floquet operator)\cite{Hannay1980,Kurlberg2000}
\begin{align}
\hat{U}(q^\prime,q)=\left(\frac{i}{K}\right)^{1/2}\exp\left[ \frac{i\pi}{K}(2q^2-2qq^\prime+2(q^\prime)^2) \right]
\label{quantum_cat}
\end{align}
where $q,q^\prime=0,1,\cdots, K-1$ label the position eigenstates. For any classical observable, one can associate a quantum observable operator $\hat O(f)$, which respects\cite{Kurlberg2000} $\hat U^\dag\hat O(f)\hat U=\hat O(f\circ M)$. 
This equation usually holds in the limit $N\to \infty$. However due to the map being linear here, it holds even at finite $N$. Therefore, for $\hat\sigma$ and $\hat \tau$, we have
\begin{equation}
\hat{U}^\dag \hat\sigma \hat U\sim \hat\sigma^2\hat\tau \qquad \hat U^\dag \hat\tau \hat U\sim \hat \sigma^3\hat\tau^2
\end{equation}
(up to some unimportant prefactor).
This result indicates that for any operator of the form $\hat{O}=\hat\sigma^q\hat\tau^p$, under unitary time evolution, it performs chaotic motion in the operator basis space spanned by $\hat{\mathcal{B}}_{mn}=\hat\sigma^m\hat\tau^n$ (with $m,n=0,1,\cdots, K-1$) which satisfies $\mbox{Tr}\hat{\mathcal{B}}_{mn}^\dag\hat{\mathcal{B}}_{m^\prime n^\prime}=K\delta_{m,m^\prime}\delta_{n,n^\prime}$. The evolution of $(q,p)$ is determined by the classical linear map defined in Eq.\eqref{classical_cat}  and gives rise to the exponential growth of the square of commutator $C(t)=\langle [\hat{O}(t),\hat O][\hat{O}(t),\hat O]^\dag\rangle$, i.e., $C(t)\sim e^{2\lambda_+t}$ when $t$ is smaller than the Lyapunov time $t_L=\log K/\lambda_+$\cite{Berenstein2015}. When $t>t_E$, the quantum correction becomes important and $C(t)$ stops to increase exponentially.

\begin{figure}[hbt]
\centering
 \subfigure[]{\label{fig:k_0_t_1} \includegraphics[width=.2\textwidth]{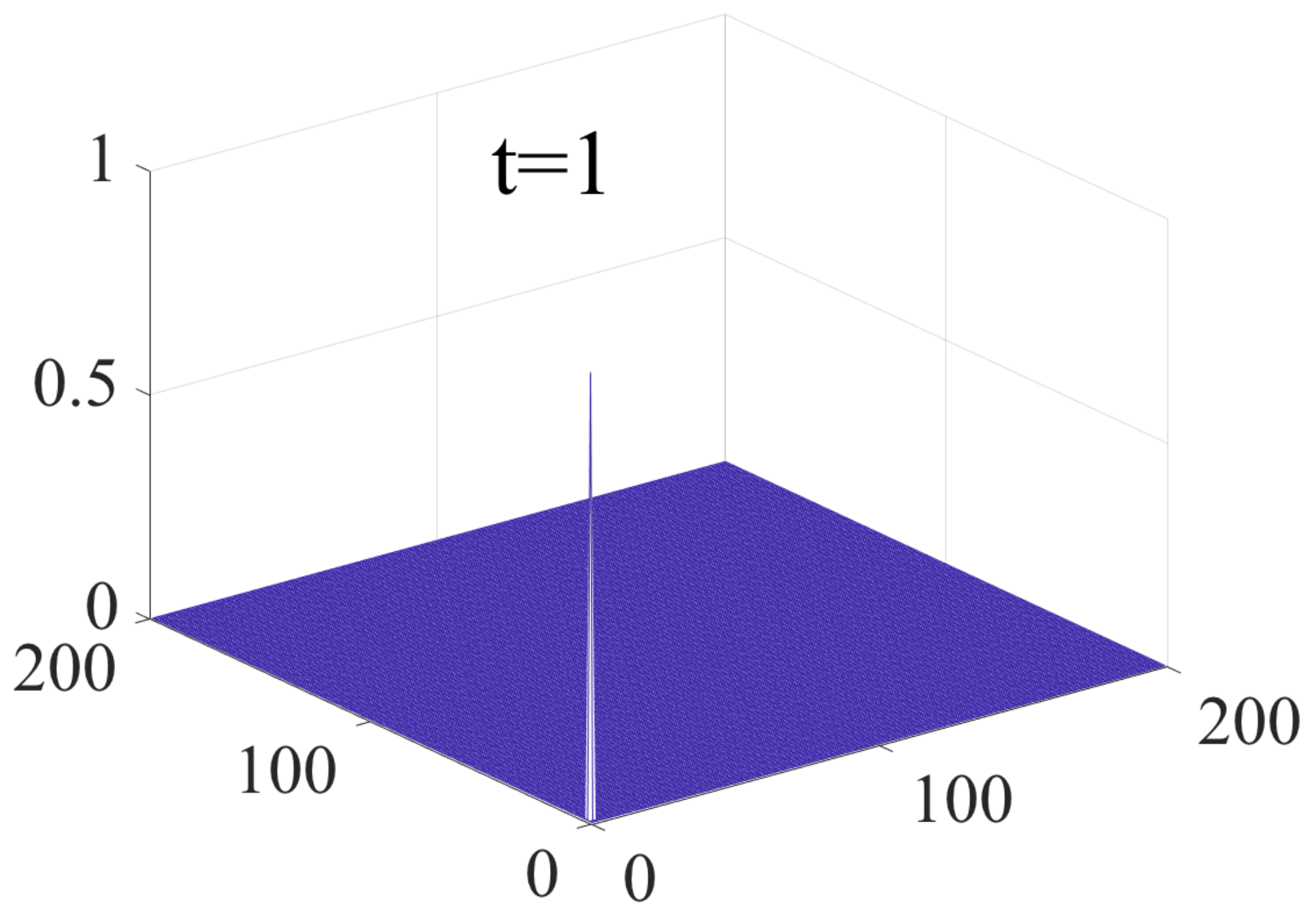}}
 \subfigure[]{\label{fig:k_0_t_4} \includegraphics[width=.2\textwidth]{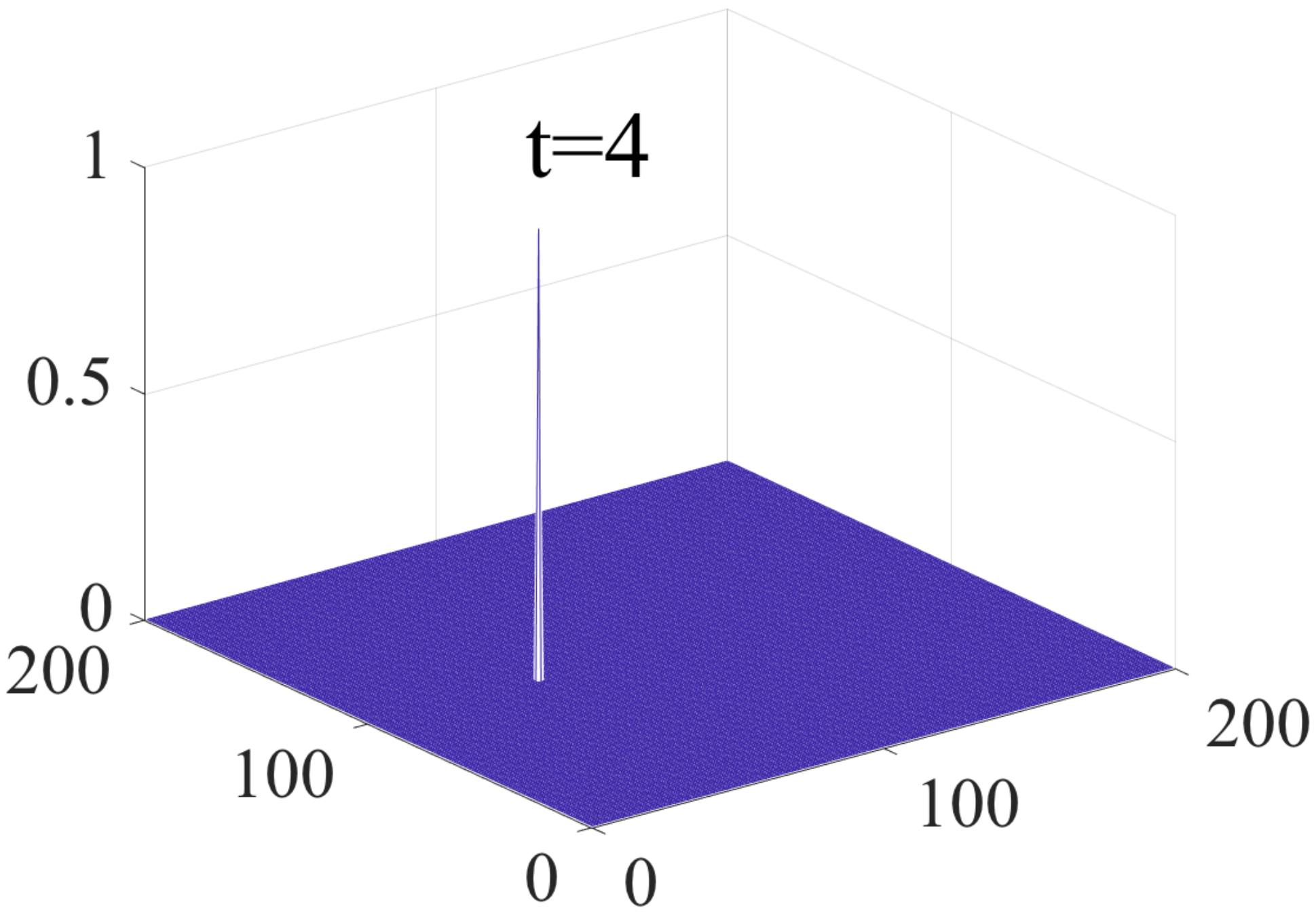}}
 \subfigure[]{\label{fig:k_0_t_7} \includegraphics[width=.2\textwidth]{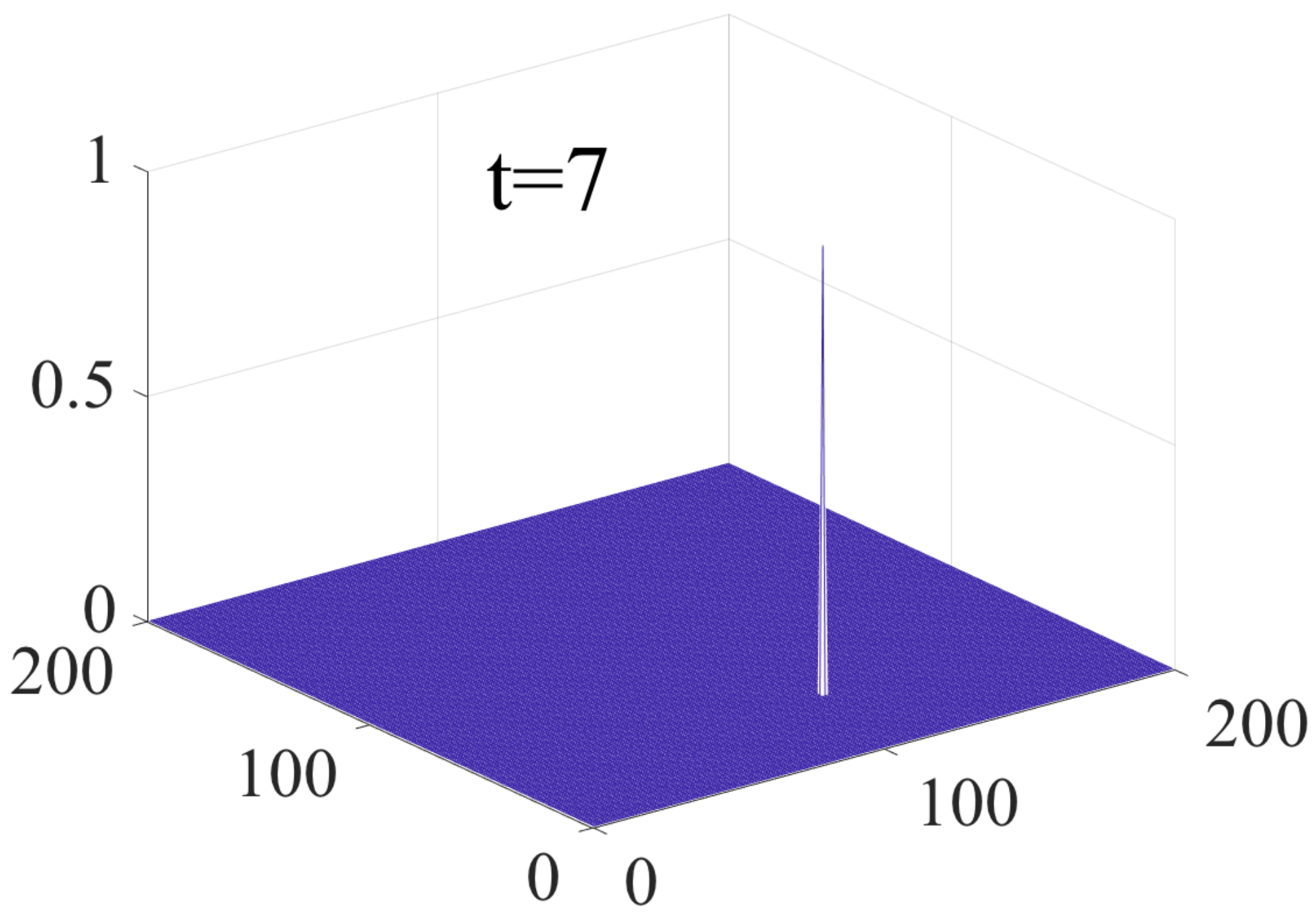}}
 \subfigure[]{\label{fig:k_0_t_11} \includegraphics[width=.2\textwidth]{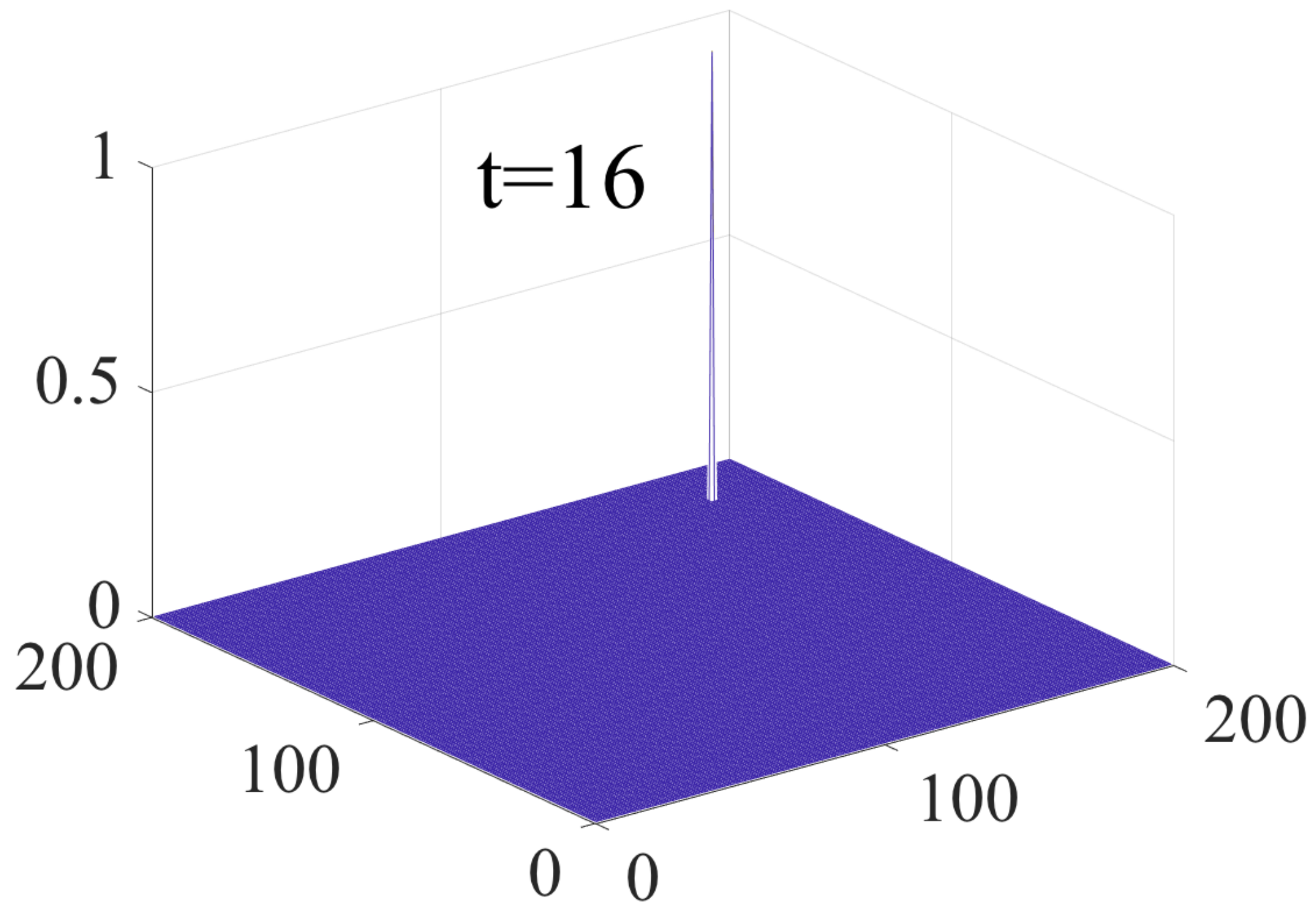}}     
  \subfigure[]{\label{fig:k_01_t_1} \includegraphics[width=.2\textwidth]{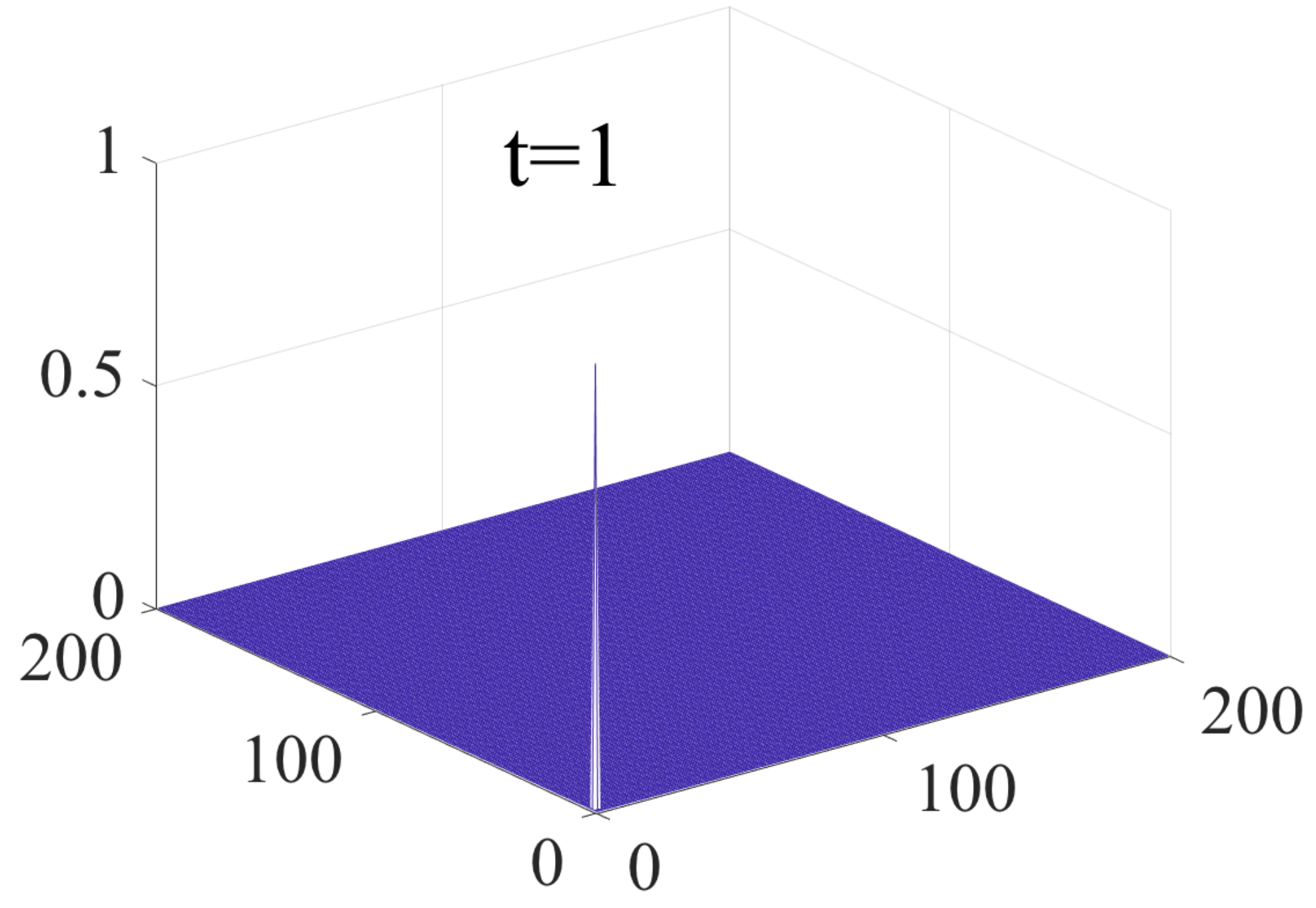}}
 \subfigure[]{\label{fig:k_01_t_4} \includegraphics[width=.2\textwidth]{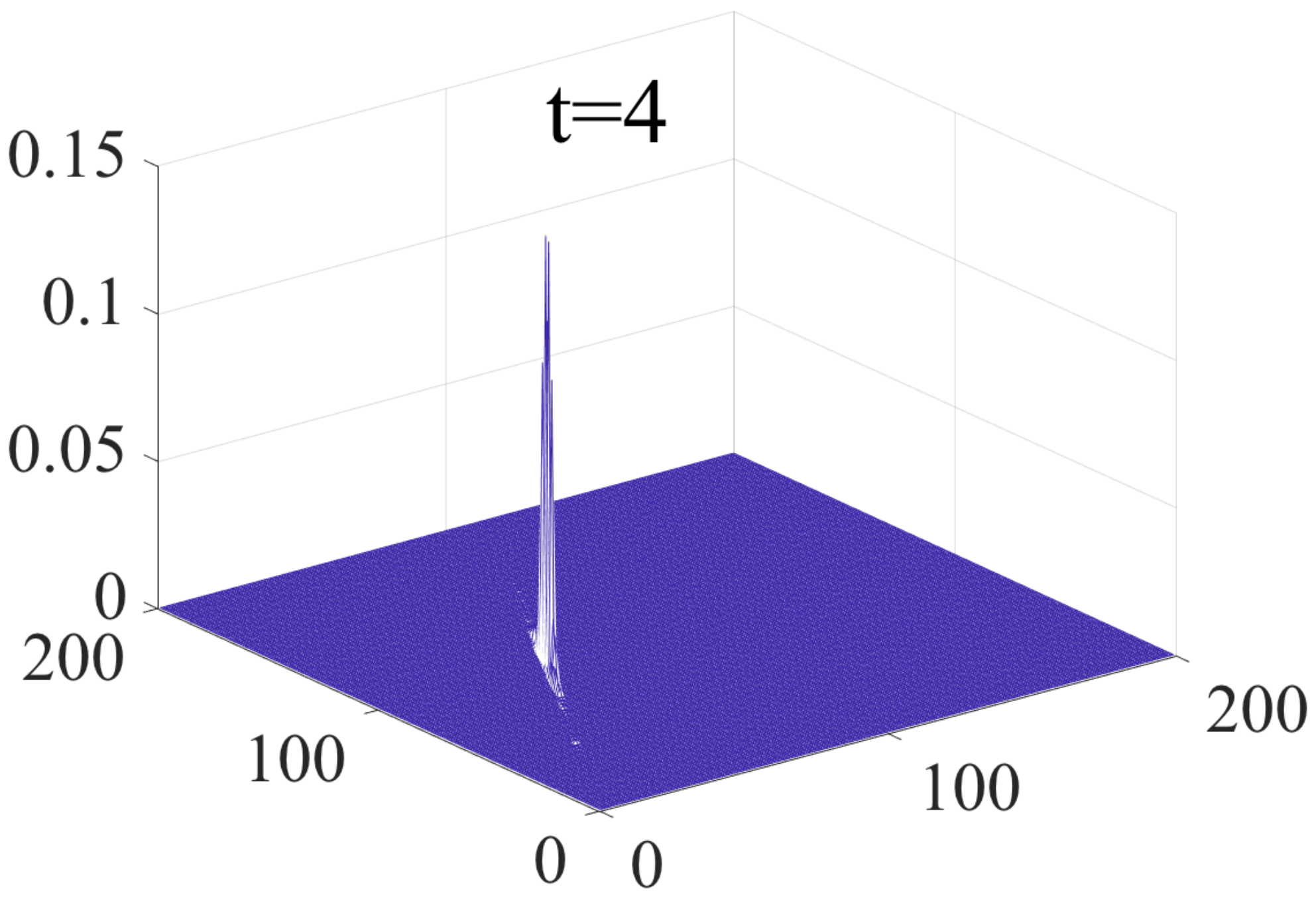}}
 \subfigure[]{\label{fig:k_01_t_7} \includegraphics[width=.2\textwidth]{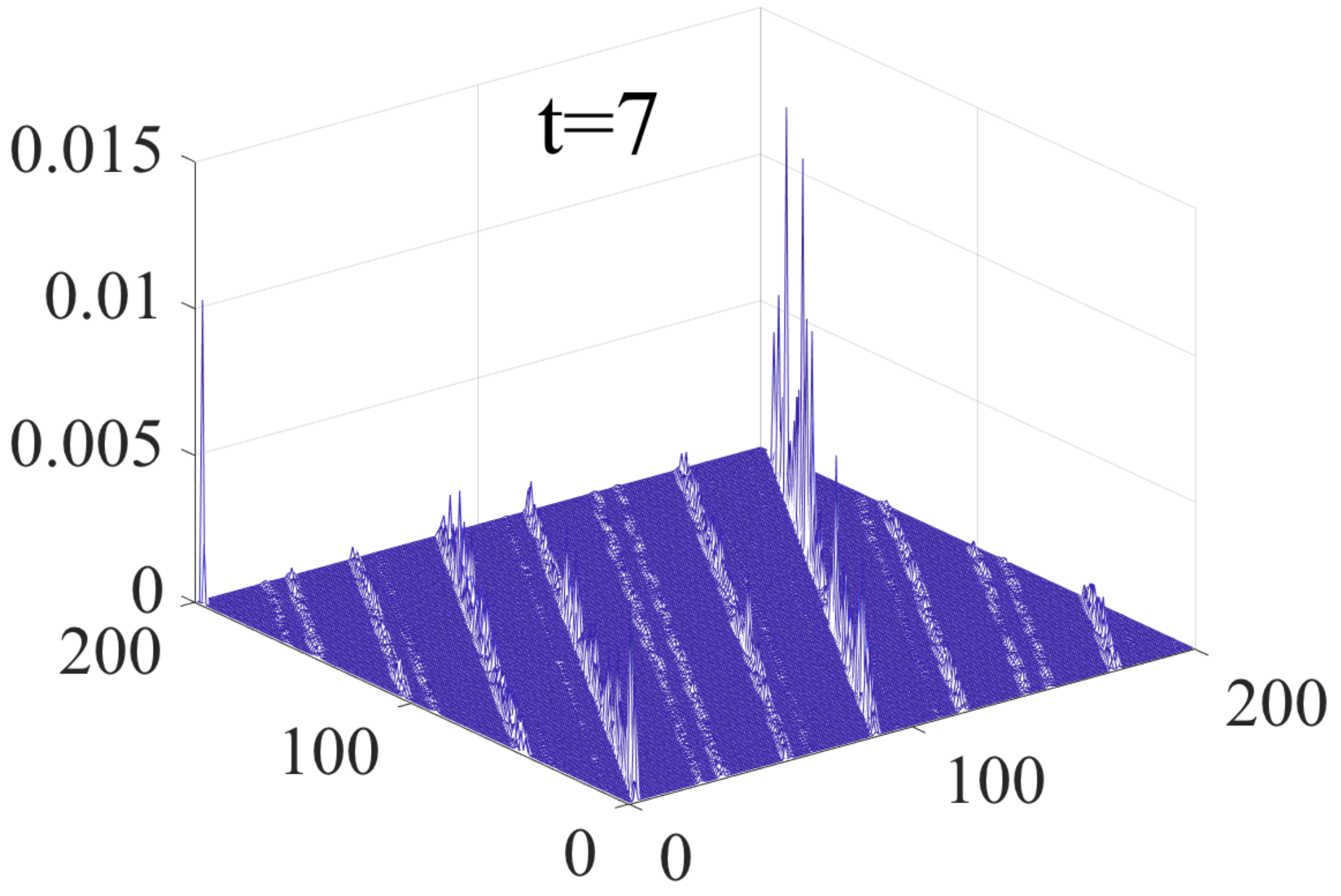}}
 \subfigure[]{\label{fig:k_01_t_11} \includegraphics[width=.2\textwidth]{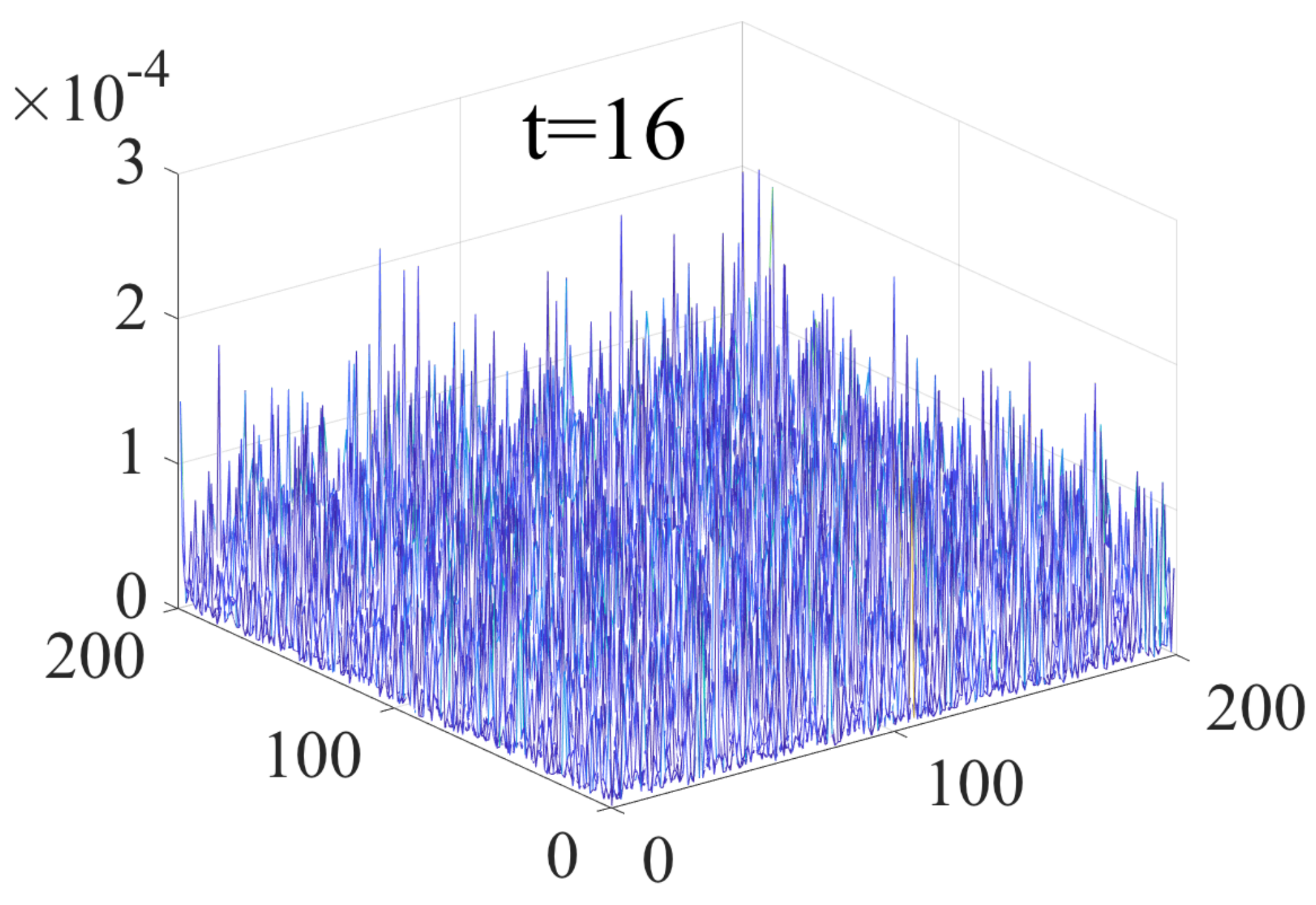}}
\caption{
The distribution of operator $\hat\sigma(t)$ on the operator basis space with $K=200$ at various time. In (a)-(d) with $\kappa=0$, the operator moves on the operator basis space determined by the classical equation of motion. In (e)-(f) with $\kappa=0.1$, the operator $\hat\sigma(t)$ spreads  out as time evolves and eventually covers the entire operator basis space.
} 
\label{fig:cat_op_dis}
\end{figure}

Notice that the exponential growth of $C(t)$ has a classical origin and is not related with the operator scrambling.  Under unitary time evolution, $\hat{O}(t)$ is always a basis operator and does not become more complicated (see the first row of Fig.~\ref{fig:cat_op_dis}). 

\begin{figure}[hbt]
\centering
 \subfigure[]{\label{fig:C_sq_cat} \includegraphics[width=.4\textwidth]{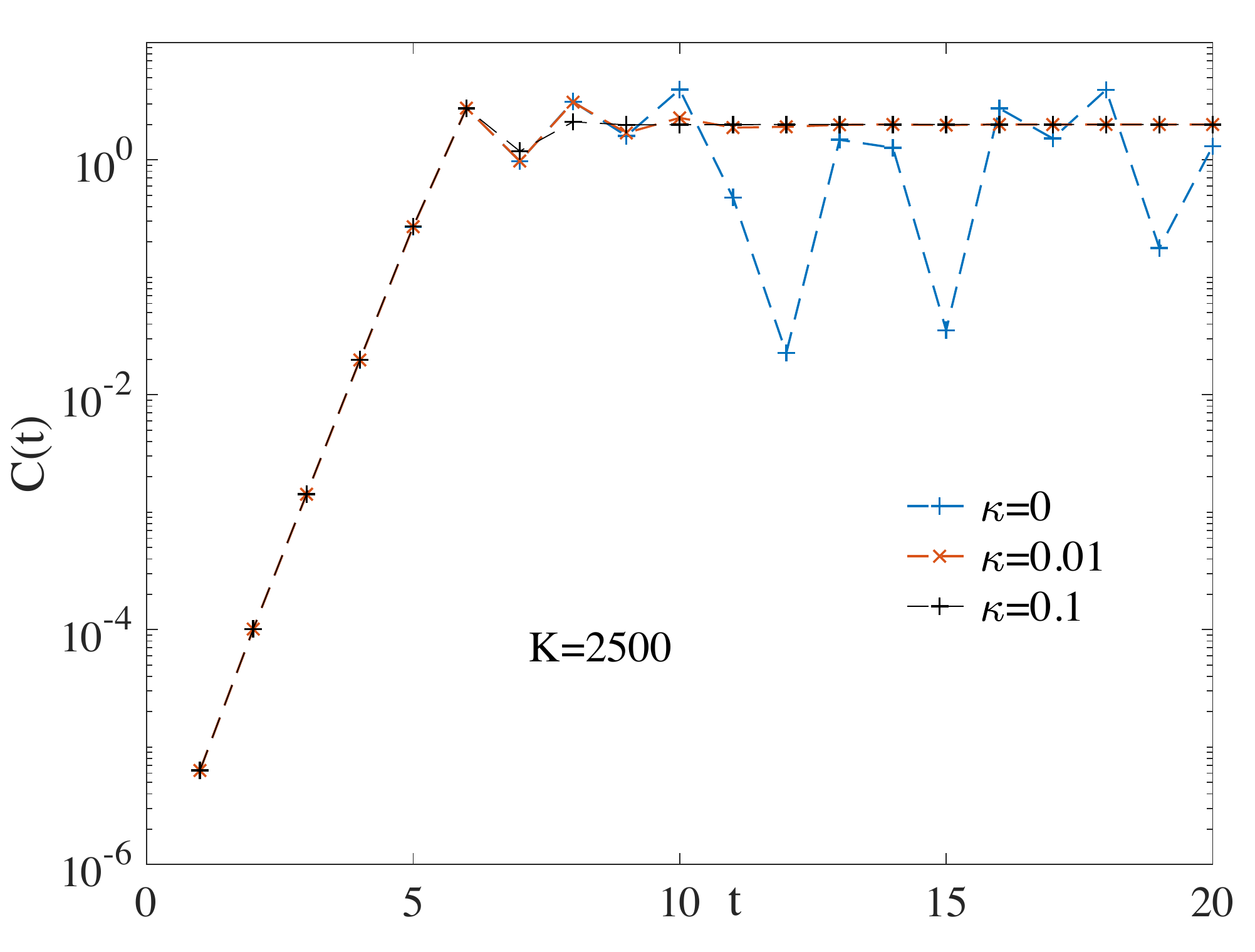}}
 \subfigure[]{\label{fig:S_op_cat} \includegraphics[width=.4\textwidth]{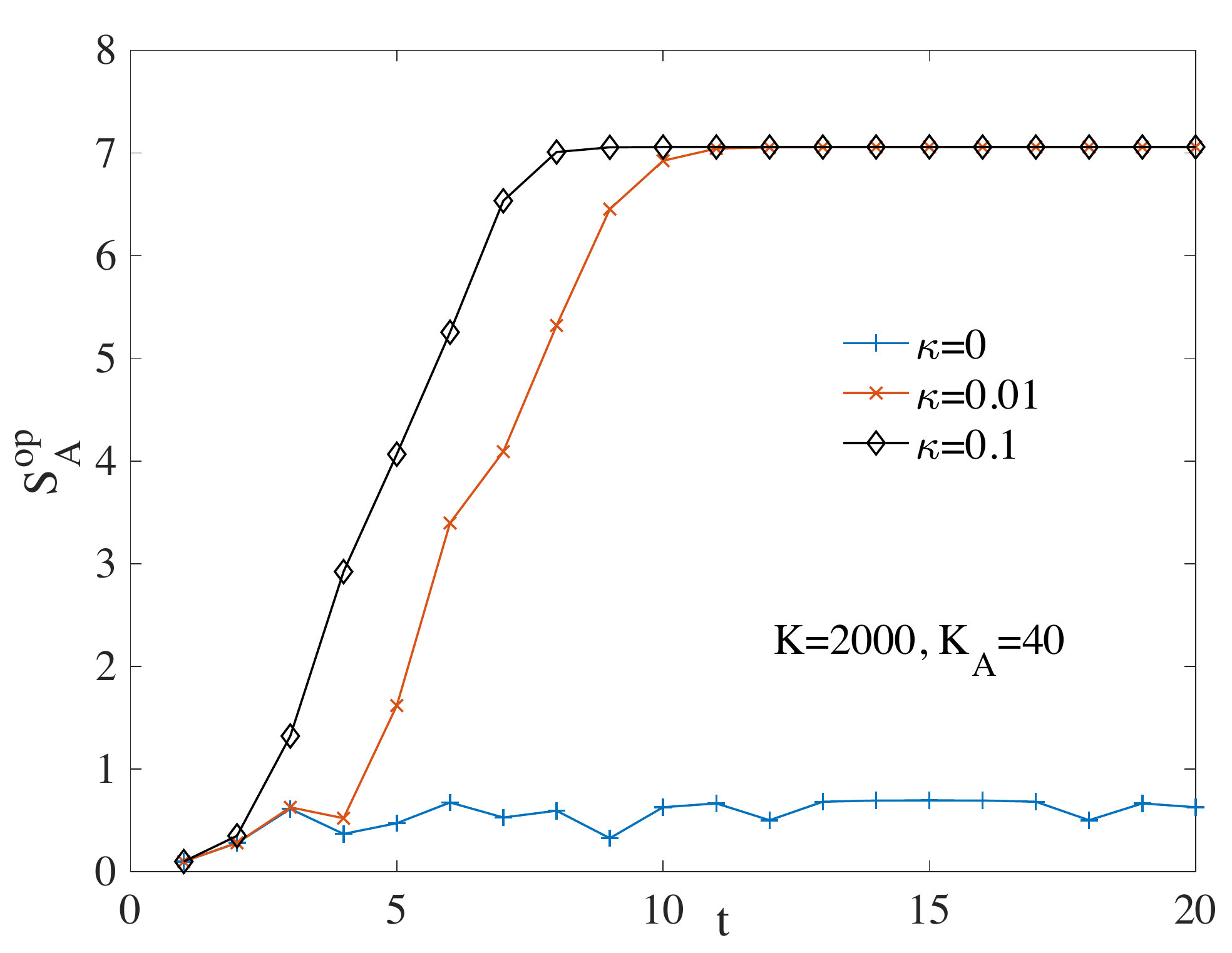}}
 \subfigure[]{\label{fig:Z_2_cat} \includegraphics[width=.4\textwidth]{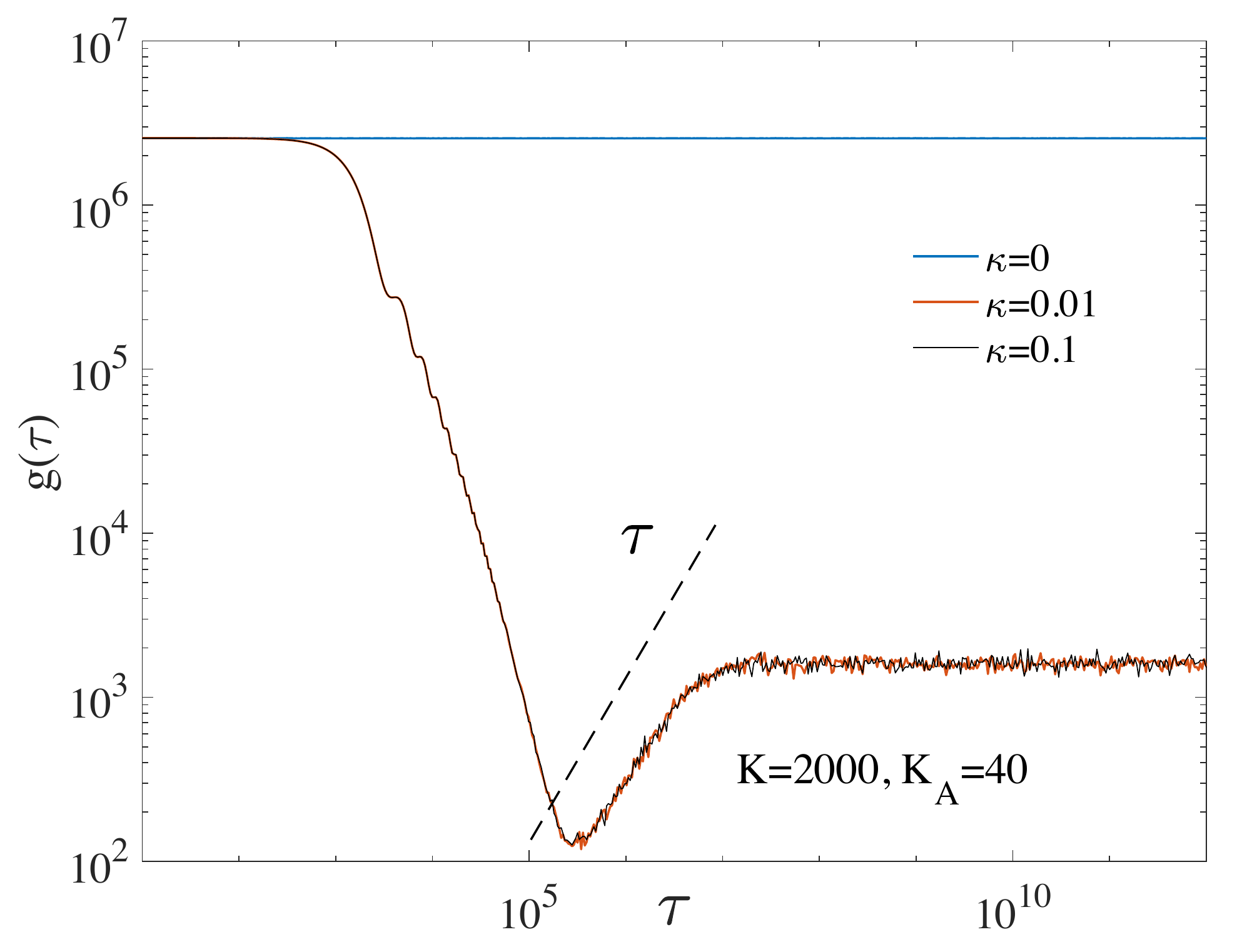}}
 \subfigure[]{\label{fig:S_op_N_cat} \includegraphics[width=.4\textwidth]{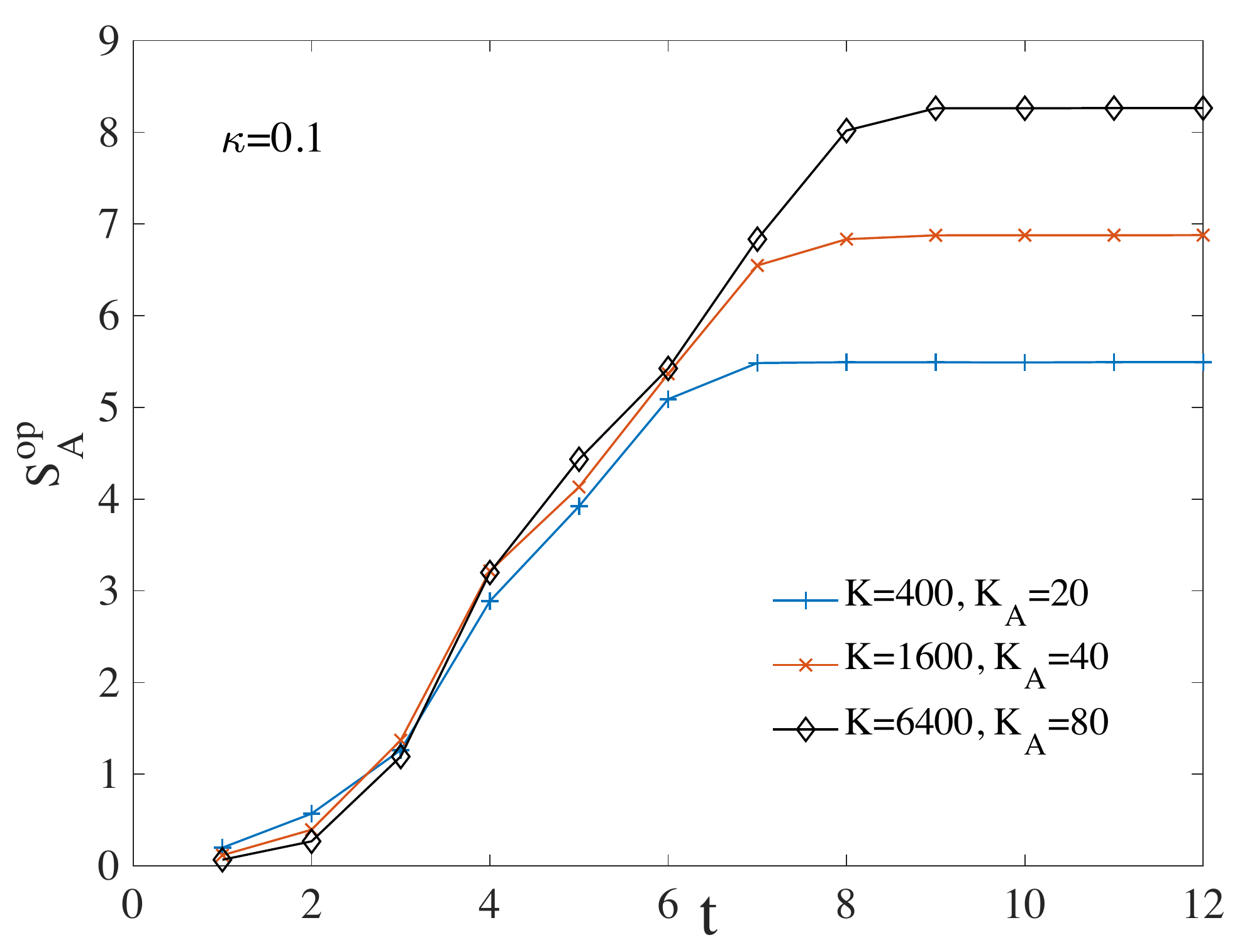}}     
\caption{
(a) The square of commutator  $C(t)=\mbox{Tr}\left\{[\hat{\sigma}(t),\hat\sigma][\hat{\sigma}(t),\hat\sigma]^\dag\right\}/K$ of the modified quantum linear map defined in Eq.\eqref{modified_cat} with $a=2, b=1, c=3$ and $d=2$. The Lyapunov exponent $\lambda_+=\log(2+\sqrt{3})$. (b) Time evolution of operator EE for $\hat\sigma(t)$ at various $\kappa$. The entire Hilbert space $K=2000$ is bipartitioned into $K_A=40$ and $K_B=50$. (c)  $g(\tau)=\langle Z(\tau)Z^*(\tau)\rangle$ for the spectrum of $\hat\rho_A^{\hat{V}}$ with $\hat V=\hat{\sigma}(t>20)$. The curves are averaged over a time-series of operator $\hat{\sigma}(t)$ in the interval $t\in[21,200]$. (d) Operator EE vs $t$ at $\kappa=0.1$ for various $K$.
} 
\label{fig:cat}
\end{figure}

To realize operator scrambling, we consider the quantum linear map perturbed by a nonlinear shear.  The new composite Floquet operator is,
\begin{align}
\hat{U}=\hat{U}_1\hat{U}_2
\label{modified_cat}
\end{align}
where $\hat{U}_2$ is the quantum linear map defined in Eq.\eqref{quantum_cat} and $\hat{U}_1$ describes a nonlinear shear\cite{Dematos1995}
\begin{align}
\langle q^\prime|\hat{U}_1|q\rangle=\exp \left[ i\frac{\kappa K}{2\pi}\left(\sin(\frac{2\pi q}{K})-\frac{1}{2}\sin(\frac{4\pi q}{K}) \right)\right]\delta_{q,q^\prime}
\end{align}
which will not have much influence on the early time dynamics as long as $\kappa$ is small. As shown in Fig.~\ref{fig:C_sq_cat}, when $\kappa\leq 0.1$, $C(t)$ always grows exponentially with the same Lyapunov exponent up to the Lyapunov time $t_L=\log K/\lambda_+\approx 6$.  Nevertheless, under the new composite Floquet operator,  we have $\hat U^\dag \hat\sigma\hat U\sim \alpha_{21}\hat\sigma^2\hat\tau+\sum_{(m,n)\neq(2,1)}\alpha_{mn}\hat\sigma^m\hat\tau^n$ and $\hat U^\dag\hat\tau\hat U\sim\beta_{32}\hat\sigma^3\hat\tau^2+\sum_{(m,n)\neq(3,2)}\beta_{mn}\hat\sigma^m\hat\tau^n$, where $\alpha_{21}$ and $\beta_{32}$ are close to one with other components being very small. As shown in the second row of Fig.~\ref{fig:cat_op_dis}, the operator mixes in the operator basis due to these small but nonzero components and can be reflected in the growth of operator EE. In Fig.~\ref{fig:S_op_cat}, we bipartition the entire Hilbert space into $A$ and $B$ subsystems and present the results of the operator EE. When $\kappa>0$, the operator EE saturates to its Page value $2\log(K_A)-K_A^2/2K_B^2$ at late time. Also the spectral correlation forms in the spectrum of $\hat\rho_A^{\hat O}(t)$. The ramp in the spectral form factor (Fig.~\ref{fig:Z_2_cat}) suggests that $\hat O(t)$ becomes a random superposition of all basis operators. This is different from the $\kappa=0$ case, where the operator EE is very small and oscillates with the time. The speed of the scrambling is determined by $\kappa$ and we always have the scrambling time $t_S\geq t_L$. Numerically, we notice that the scrambling time is also proportional to $f(\kappa)\log K$ with the prefactor $f(\kappa)$ as a function of parameter $\kappa$ (Fig.~\ref{fig:S_op_N_cat}).

The modified quantum linear map represents a large class of quantized model which is chaotic in the classical limit. In these models, the time evolution of the quantum observable operator involves both classical and quantum parts:  (1) it follows the classical motion in the operator ``phase space" and (2) in the meanwhile it evolves into a superposition of basis operators and eventually becomes a highly entangled random operator. The first part is determined by the classical chaotic dynamics while the second part implies operator scrambling which is {\it unique in quantum chaotic system} and contributes to the growth of operator EE. As time evolves, the operator loses the memory of the initial state, accompanied by the formation of universal spectral correlation in the spectrum of the operator reduced density matrix of the subsystem. All together, these features give a definition of quantum chaos and our first and second example in Sec.~\ref{sec:local_int} and Sec.~\ref{sec:2-local} indeed satisfy this criterion (although they don't have semiclassical limit).




\section{Discussion and Conclusion}
\label{sec:conclusion}

In conclusion, we explore the scrambling of a simple operator in three representative examples and discuss its importance in quantum chaos. In the chaotic spin-$1/2$ chain, we illustrate that a local operator can expand linearly in time. It therefore has the scrambling time linearly proportional to the system size. In the 2-local Hamiltonian model, we give an intuitive picture in terms of the operator height to demonstrate that the scrambling is much faster than the first example: its scrambling time scales as $\log N$, where $N$ is the number of total spin. In both models, we use the operator entanglement entropy and spectral correlation in the spectrum of the operator reduced density matrix to characterize quantum chaos and the emergence of the random matrix physics. In the quantum linear map, although the square of the commutator $C(t)$ can grow exponentially with the time, we find that the quantum  operator does not scramble at all. The operator scrambling can occur once we make some modification in the Floquet operator. This suggests that the quantum chaos is not always associated with the exponential growth of the square of the commutator $C(t)$ and one should look at quantum scrambling instead. 

Here, we list several possible directions to explore in the future. First, the operator scrambling we discuss in this paper is indeed at infinite temperature. We can generalize to finite temperature with the thermal operator $e^{-\beta \hat H/2}\hat V(t)e^{-\beta\hat H/2}$ to study the influence of temperature on scrambling. Secondly, the quantum dot model we investigate in Sec.~\ref{sec:2-local} can be used as building blocks for one dimensional lattice model with large onsite Hilbert space. We plan to investigate operator scrambling in this model, compare the result with possible solutions proposed in Sec.~\ref{dot_chain} and explore the crossover to  the behaviors of spin-$1/2$ chain with small onsite Hilbert space. Thirdly, the modified quantum linear map we study in Sec.~\ref{sec:linear_map} is one of the quantum  mechanical models that has classical chaotic limit. It would be interesting to have a better understanding of operator scrambling in these models in the semiclassical limit.

\acknowledgements We acknowledge  Eduardo Fradkin, Chunxiao Liu, Andreas W.W. Ludwig, Chetan Nayak, Adam Nahum and Cenke Xu for useful discussion and previous collaborations on related topic. XC was
supported by a postdoctoral fellowship from the Gordon and Betty
Moore Foundation, under the EPiQS initiative, Grant GBMF4304, at
the Kavli Institute for Theoretical Physics. 
We acknowledge support from the Center for Scientific Computing from the CNSI, MRL: an NSF MRSEC (DMR-1121053).

\bibliographystyle{unsrt}
\bibliography{chaos_scrambling,op_scrambling}

\end{document}